\documentstyle[12pt,epsf]{article}

\begin{document}
\begin{flushright}
SLAC-PUB-8459\\
May 2000
\end{flushright}
\vfill
\begin{center}
{\LARGE {Status of PEP-II and BaBar}
\footnote{\baselineskip=13pt Work supported by the Department
of Energy, contract DE--AC03--76SF00515.}}

\vspace{15mm}
{\bf J. Dorfan}\\
\vspace{5mm}
{\em Stanford Linear Accelerator Center, \\
Stanford University, Stanford, California 94309}

\end{center}

\vfill

\begin{center}
Invited talk presented at \\
19th International Symposium on Lepton and Photon Interactions at High Energies\\
Stanford University, Stanford, California \\
August 9--14, 1999
\end{center}
\vfill
\newpage

\section{Introduction}

The SLAC B Factory 
\index{subject}{asymmetric $B$-factories}
was approved by President Clinton in October 1993.
\index{subject}{B-factories}
The inaugural meeting of the detector collaboration was held at SLAC in
December 1993, at which time the 9 nation BaBar collaboration began.
First funding for the machine construction was released in January 1994,
launching the SLAC/LBNL/LLNL PEP-II project collaboration \cite{ref1}.  The
detector technical design report was completed in March 1995 \cite{ref2}.

PEP-II was completed in July 1998 and first collisions were observed on
July 22, 1998.  Commissioning without BaBar concluded in February 1999,
at which time the BaBar detector commissioning ceased and BaBar was
moved from the off-beamline to the on-beamline position.  Running
commenced in early May 1999, and first collisions with BaBar were
observed on May 26, 1999.

In this report, the status of the PEP-II machine and the BaBar detector
are presented.  Performance will be presented as of the Lepton Photon
Conference (August 1999) with subsequent performance levels footnoted as
appropriate.  This report is made on behalf of the SLAC/LBNL/LLNL
team that built and commissioned PEP-II, the SLAC Accelerator Department
that commissioned and operate the machine, and the 71 institution, 9
nation detector collaboration that built, commissioned, and operate
BaBar.  

\section{PEP-II Machine}

\begin{figure}[htbp]
\begin{center}
{\hspace{1in}\epsfbox{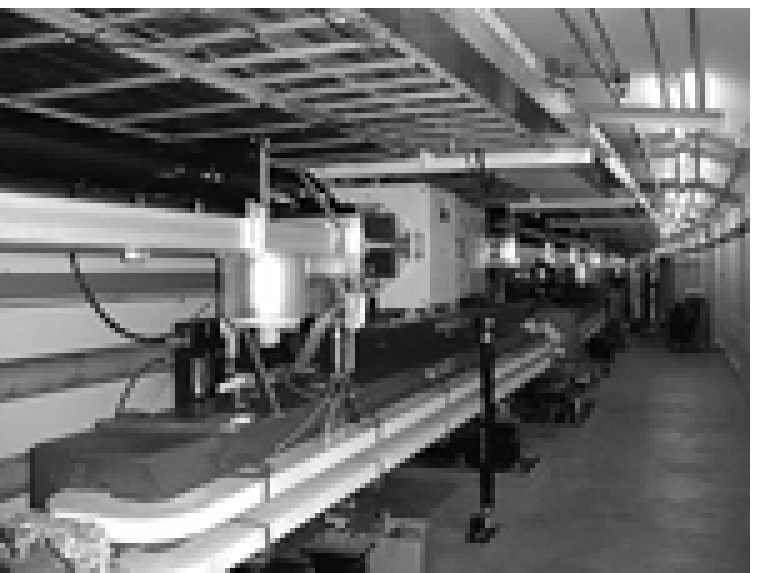}}
\caption[]{View of the two PEP-II storage rings.}
\label{dfig:1}
\end{center}
\hfill
\begin{center}
{\hspace{1in}\epsfbox{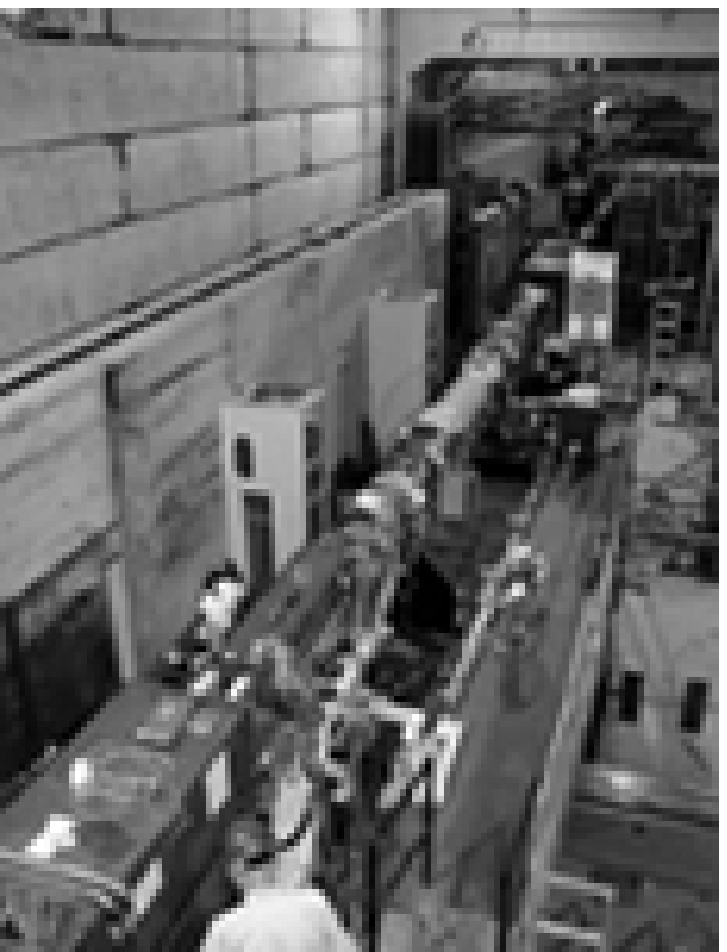}}
\caption[]{The PEP-II interaction region, before roll-in of the BaBar detector.}
\label{dfig:2}
\end{center}
\end{figure}

 The PEP-II \index{subject}{PEP-II} design parameters are shown in Table~\ref{dtab:1d}.
The machine consists of two storage rings, vertically stacked---the
high-energy (9 GeV) ring stores electrons and the low energy (3 GeV)
ring stores positrons (see Fig.~\ref{dfig:1}).
Collisions occur in a single interaction region (IR) (see Fig.~\ref{dfig:2}), at
which point the upper low energy ring (LER) has been brought down into
the plane of the high energy ring (HER).

\begin{table}[htb]
\begin{center}
\begin{tabular}{|c|c|c|}
\hline
& $e^+$ & $e^-$ \\ \hline
CM energy (GeV) & \multicolumn{2}{c|}{10.580} \\ \hline
Beam energy (GeV) & 3.119 & 8.973 \\ \hline
Beam current (A)     & 2.15   & 0.75 \\ \hline
$\beta_x^* \ |\ \beta_y^*$  (cm) & 50 $|$ 1.5 & 50 $|$ 1.5 \\ \hline
$\epsilon_x\  |\ \epsilon_y$ (nm)   & 49 $|$ 1.5  & 49 $|$ 1.5 \\ \hline
$\sigma_x^*$ ($\mu$ m) & \multicolumn{2}{c|}{157} \\ \hline
$\sigma_y^*$ ($\mu$ m) & \multicolumn{2}{c|}{4.7} \\ \hline
$\sigma_z$  (mm) & 12.3  & 11.5 \\ \hline
Luminosity & \multicolumn{2}{c|}{$3 \times 10^{33} {\rm  cm}^{-2} {\rm s}^{-1}$}
\\ \hline
Tune shift & \multicolumn{2}{c|}{0.03} \\ \hline
Beam aspect ratio (v/h at IP) & \multicolumn{2}{c|}{0.03} \\ \hline
Number of colliding bunches & \multicolumn{2}{c|}{1658} \\ \hline
Bunch spacing (m) & \multicolumn{2}{c|}{1.26} \\ \hline
Beam crossing angle & \multicolumn{2}{c|}{0 (head-on)} \\ \hline
\end{tabular}
\caption[]{PEP-II design parameters.}
\label{dtab:1d}
\end{center}
\end{table}

Collisions are head-on; beam separation is done magnetically.  Electrons
and positrons are supplied by the SLAC linac.  Extraction and injection
is done at the collision energy, so that no ramping of the stored beam is
required.  The RF system components are identical for the two rings---1.2 
MW CW klystrons and warm copper cavities, which are actively damped.
The on-cavity damping is not sufficient to ensure stable beams at high
currents.  Accordingly, bunch-by-bunch damping is supplied in both the
transverse and longitudinal planes.  The damping is primarily needed to
ameliorate the effects of multi-bunch instabilities.  A much more
detailed description of PEP-II can be found in \cite{ref1}.

The machine group took on aggressive goals with respect to early peak
and integrated luminosity performance.  To achieve these goals,
machine commissioning was phased in time.  The high energy and low
energy injection systems were completed early.  The 
HER was completed in July 1997, allowing commissioning of the
HER in 1997 and also testing of many systems that were common to the
LER.  The full machine was completed in July 1998 with the LER and 
IR being the last systems to finish.  However,
injection into a partial LER was achieved in early 1998.  The phased
approach to commissioning was an effective mechanism for early success
with the full machine.  Two weeks after completion of the full machine, the
first collisions were observed.

\section{BaBar Detector}

\begin{table}[htb]
\begin{center}
\vspace{.2in}
\begin{tabular}{|l|l|l|l|}
\hline
{\bf Det.} & {\bf Technology} & {\bf Dimensions} & {\bf Performance} \\ \hline
SVT 
& Double-sided 
& 5 layers 
& $\sigma_z=\sigma_{xy} = 50\ \mu {\rm m}/p$ \\ 
&  silicon strips 
& $r = 3.2\ {\rm  to}\  14.4\ {\rm cm}$  
& $ \qquad \oplus 15\mu{\rm m}\ @\ 90^\circ$ \\ 
& & $-0.87 <$ cos$\theta$ $<$ 0.96 
&  $\sigma_\emptyset = \sigma_\theta = 1.6 \ {\rm mr/p}\ @\ 90^\circ$\\[1ex] \hline
DC 
& Drift chamber 
& 40 layers 
& $\sigma_{pt}/p_t = [0.21\% + 0.14\%\times p_t]$ \\ 
& & r = 22.5 to 80 cm &  \\ 
& & -111 $<$ z $<$ 166 cm & \\[1ex] \hline
PID 
& DIRC 
& $ 1.75 \times 3.5\ {\rm  cm}^2$  quartz 
& $ N_{pe} = 20 - 50$ \\ 
& & $-0.84 < \cos\theta < 0.90$ 
& \parbox[t]{1.92in}{$\geq 4 \sigma K/\pi$ separation for all B decay products} \\[3ex] \hline
CAL
& CsI(Tl)
& 16 to 17.5 $X_0$ 
& $\sigma_E/E= [1\%/E\ {\rm (GeV)}]^{1/4}$ \\ 
& & $\sim 4.8 \times 4.8$  cm crystals
&  $ \qquad \oplus 1.2\%$ \\
& & &  $\sigma_\theta = 3\mu\rho/\sqrt{E\ k{\rm (GeV)}}\ \oplus $ 2 mr \\[1ex] \hline       
MAG
& Superconducting & IR = 1.40 m & B = 1.5 T \\
& segmented steel & L = 3.85 m
& \\[1ex] \hline
IFR
& RPC
& 
18-19 planar layers
& $E_\mu  > 90\% $ \\
& & $+$ 4 cylindrical layers
& for $ P_\mu >$ 0.8 GeV/c \\[1ex] \hline
\end{tabular}
\caption[]{Performance specifications for the various components of the
BaBar detector.}
\label{dtab:2d}
\end{center}
\end{table}

\begin{figure}[htbp]
\begin{center}
{\hspace{1in}\epsfbox{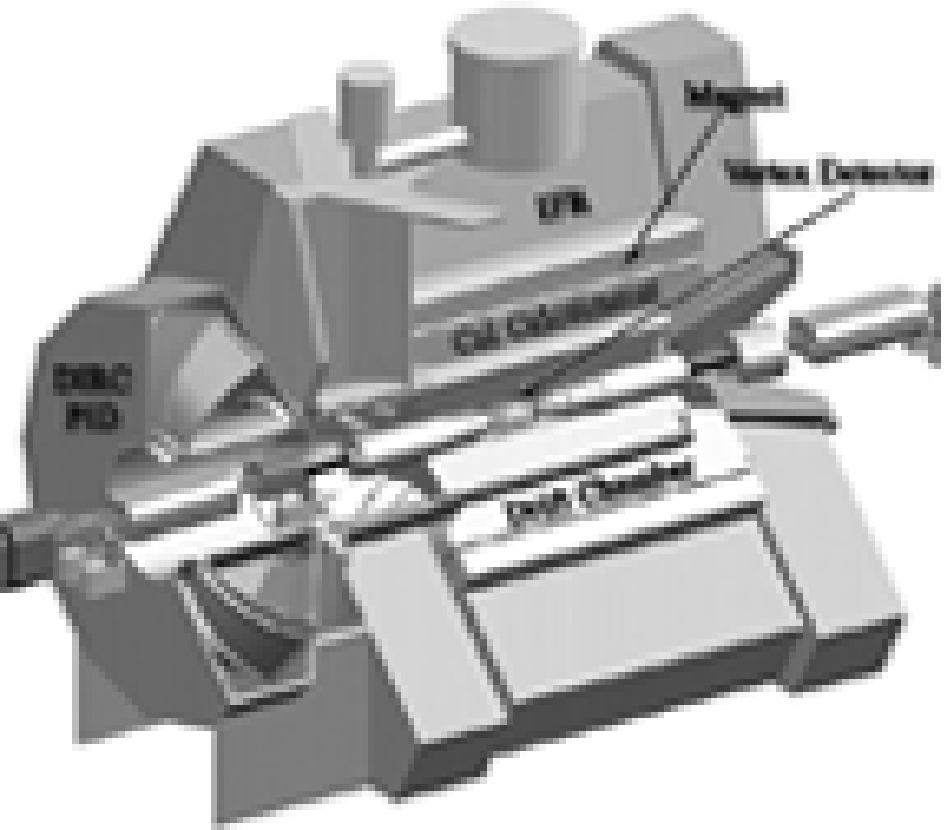}}
\caption[]{Cutaway view of the BaBar detector.}
\label{dfig:3}
\end{center}
\vfill
\begin{center}
{\hspace{1in}\epsfbox{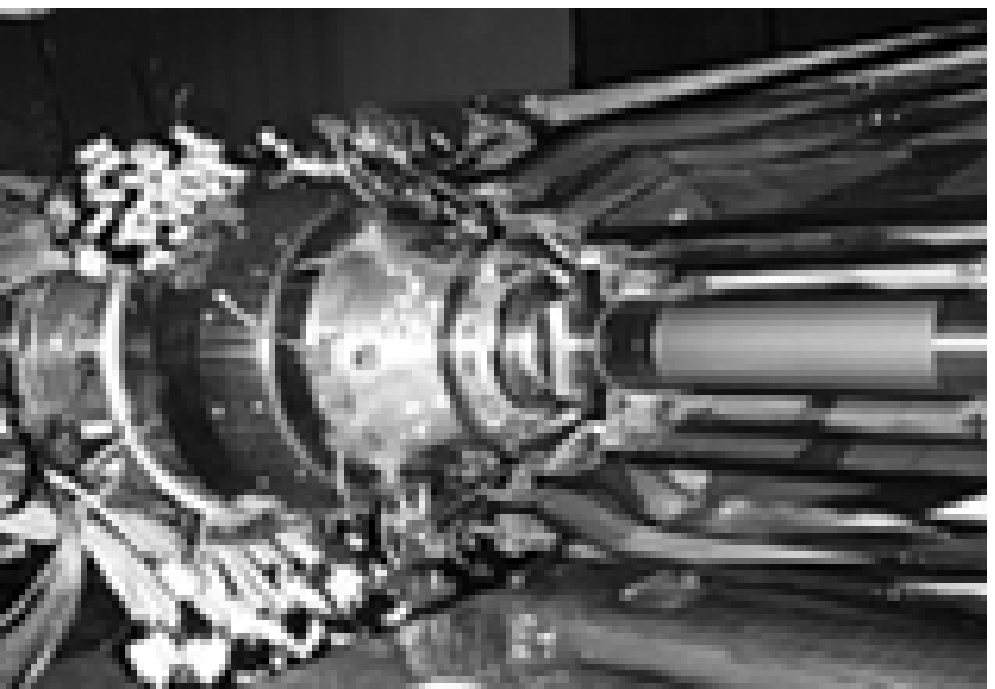}}
\caption[]{The BaBar silicon strip vertex detector.}
\label{dfig:4}
\end{center}
\end{figure}

The BaBar \index{subject}{BaBar detector}
design performance specifications are shown in Table~\ref{dtab:2d}. An
isometric view of the detector is shown in Fig.~\ref{dfig:3}, indicating the
\index{subject}{silicon vertex detector}
silicon vertex detector, drift chamber, Cherenkov radiation particle
identification system, cesium iodide calorimeter, superconducting 1.5
Tesla magnet, and instrumented flux return \cite{ref2}.  The accelerator magnets
in the IR come within 20 cm of the collision point.  In fact, the vertex
detector is actually mounted on the final accelerator magnets.

\begin{figure}[htbp]
\begin{center}
{\hspace{1in}\epsfbox{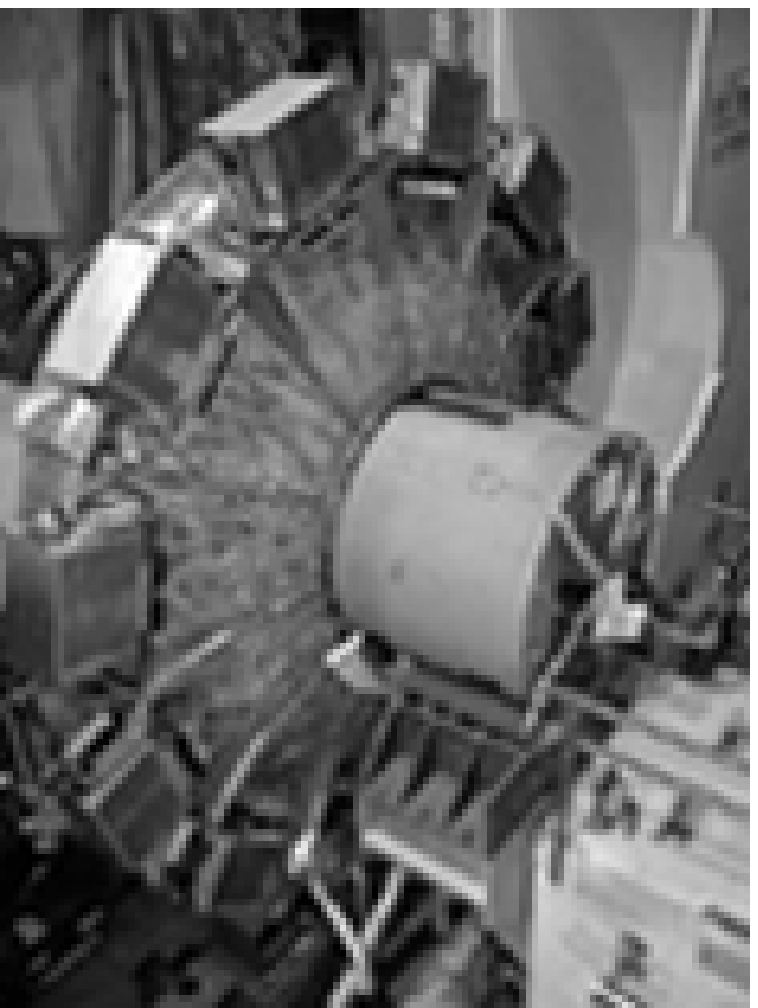}}
\caption[]{Phototube array for the BaBar DIRC particle identification system.}
\label{dfig:5}
\end{center}
\vfill
\begin{center}
{\hspace{1in}\epsfbox{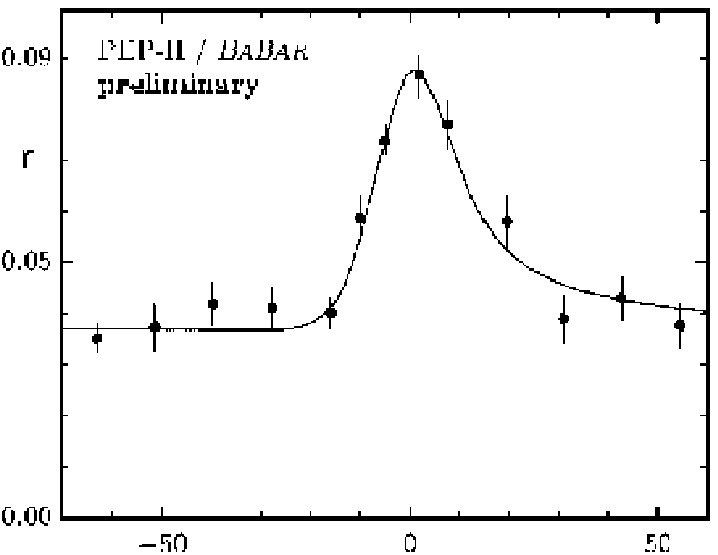}}
\caption[]{BaBar scan of the $\Upsilon(4s)$ resonance.}
\label{dfig:6}
\end{center}
\end{figure}

The vertex detector, one-half of which is shown in Fig.~\ref{dfig:4}, is
comprised of 5 layers of silicon strip detectors read out with
radiation-hard electronics.  The drift chamber is a conventional, small
cell design with 8 layers of axial wires interleaved with 4 layers of $\pm$ 3$^0$
stereo wires.  A low mass helium-based gas reduces multiple scattering.
Surrounding the drift chamber is the DIRC \index{subject}{Cherenkov detector}
system, 48 quartz bars running azimuthally.
Cherenkov light generated in the quartz bars is propagated along the
bars, exiting at the rear end of the detector where the exit angle is
mapped by an array of 11,000 phototubes (see Fig.~\ref{dfig:5}).
The pattern of phototube hits can be used to measure the Cherenkov angle
and thereby identify the radiating particle.  Photons and
electrons are identified in the \index{subject}{cesium diodide calorimeter}
cesium iodide calorimeter.  Because of
the asymmetry of the machine, there is no need for electromagnetic
calorimeter coverage in the backward region.  The steel superstructure
used to return the magnetic flux is made in layers interspersed with
\index{subject}{resistive plate chamers}
resistive plate chambers.  This instrumented flux return is used for
muon and $K_L^0$ identification.  BaBar has chosen a pioneering approach to
data acquisition and data analysis software.  All software (both the coding and the databases)
uses the object oriented approach  The
accelerator uses photons from the process $e^+e^- \rightarrow e^+e^-\gamma$ to derive a fast
luminosity signal.  However, the Bhabha rate is high enough at moderate
luminosities to provide an online luminosity measurement from the
detector.  Luminosities quoted here come from the Bhabhas in BaBar.

\section{Machine and detector performance}

\begin{figure}[htbp]
\begin{center}
{\hspace{1in}\epsfbox{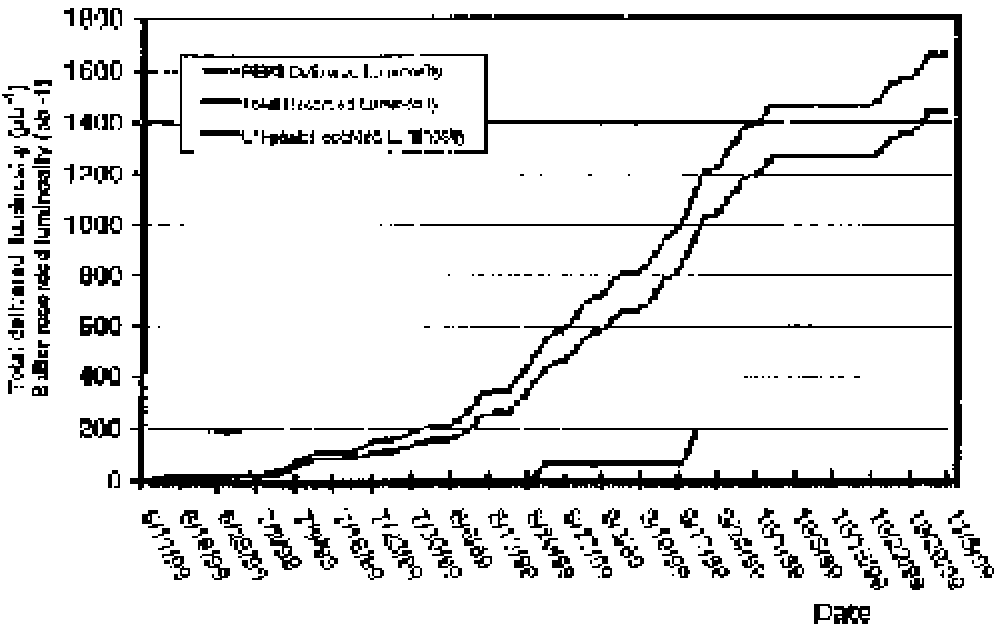}}
\caption[]{Integrated luminosity accumulated by PEP-II as a function of time.}
\label{dfig:7}
\end{center}
\vfill
\begin{center}
{\hspace{1in}\epsfbox{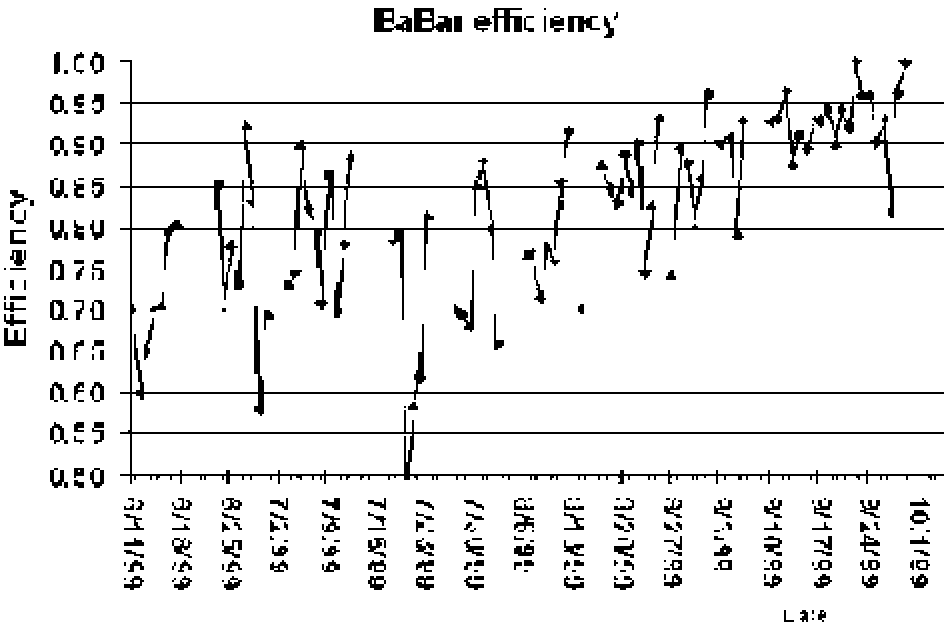}}
\caption[]{BaBar's efficiency of data logging, as a function of time.}
\label{dfig:8}
\end{center}
\end{figure}

The month of June 1999 was used to integrate the machine and detector
programs into a system which could run steadily for physics.  The
luminosity was improved and machine related backgrounds were understood.
The \index{subject}{Upsilon (4s) resonance}
$\Upsilon(4s)$ resonance was mapped out in a 3-day run as shown in Fig~\ref{dfig:6}.
Using about 1/3 of the data, resonance parameters of $M(\Upsilon_{4s}) = 10.5841 \pm
0.0007$  GeV and $\Gamma (\Upsilon_{4s}) = 11.1 \pm 3.4$ MeV were obtained.  In July
1999, persistent physics running began, interspersed with prolonged periods
for machine development.  Performance results for the machine  as of the time of LP99 
are given in Table~\ref{tab:3}.  In the table, typical running parameters with BaBar 
are contrasted with the best that
was achieved during commissioning without BaBar.  As of the end of July
1999, BaBar had logged about 200 pb$^{-1}$, as can be seen from 
Fig.~\ref{dfig:7}.\footnote{ 
As of the end of November 1999, BaBar had logged 2 fb$^{-1}$, and the record
peak luminosity was $1.43 \times 10^{+33}$  cm$^{-2}$ sec$^{-1}$.}

\begin{table}[htbp]
\begin{center}
\begin{tabular}{|l|c|c|c|c|}
\hline \hline
\multicolumn{5}{|c|}{PEP-II HER Performance Results} \\ \hline\hline
{\bf Parameter} 
& {\bf Units} &  {\bf Design} & {\bf Commissioning} 
& {\bf Running}  \\[1ex]  \hline \hline
Energy
& GeV
& 9.0
& 9.0, ramp to & 9.0, ramp to\\[-1ex] 
& & &  9.1 \& bk
&  8.84-9.04 \\[1ex] \hline 
Single bunch current
& mA
& 0.6
& 12
& 0.55 \\[1ex] \hline 
Number of bunches
&  & 1658
& 1658
& 415 \\[1ex] \hline
Total beam current
& A
& 0.995
& 0.75
& 0.25 \\[1ex] \hline
Beam Lifetime
& hours
& 4
& 8 hrs & 6 hrs @  \\[-1ex] 
& & & @ 250 mA & 250 $\mu$amp \\[1ex] \hline
Max. Injection Rate
& mA/sec
& 2.1 @ 60 Hz
& 2.5 @ 10 Hz
& 0.6 @ 10 Hz \\[1ex] \hline \hline
\multicolumn{5}{|c|}{PEP-II LER Performance Results} \\ \hline \hline
Energy
& GeV
& 3.1
& 3.1
& 3.1 \\[1ex] \hline
Single bunch charge
& mA
& 1.3
& 7.0
& 1.9 \\[1ex] \hline
Number of bunches
& & 1658
& 1658
& 415 \\[1ex] \hline
Total charge
& A
& 2.14
& 1.171
& 0.8 \\[1ex] \hline
Beam Lifetime
& hours
& 4
& 50 min & 120 min  \\[-1ex] 
& & & @ 800 mA
&@ 800 mA \\[1ex] \hline
Max. Injection Rate
& mA/sec
& 5.9 @ 60 Hz
& 3.0 @ 10 Hz
& 2.7 @ 10 Hz \\[2ex] \hline \hline
\multicolumn{5}{|c|}{PEP-II Collider Performance Results} \\ \hline \hline
Luminosity
& cm$^{-2}$  sec$^{-1}$
& $3 \times 10^{33}$
& $5.2 \times 10^{32}$
& $5.6 \times 10^{32}$ \\[1ex] \hline
Specific Luminosity
& cm$^{-2}$  sec$^{-1}$  mA$^{-2}$
& $3.1 \times 10^{30}$
& $1.7 \times 10^{30}$
& $2.1 \times 10^{30}$ \\[1ex] \hline
Horizontal Spot Size
& $\mu$m
& 220
& 220
& 220 \\[1ex] \hline 
Vertical Spot Size
& $\mu$m
& 6.6
& 8.6
& 10.8\\[1ex]  \hline\hline
\end{tabular}
\caption[]{Performance of PEP-II.  ``Commissioning" refers to the best results
obtained without the detector in place.  ``Running" refers to typical results with
BaBar.}
\label{tab:3}
\end{center}
\end{table}

Machine related backgrounds were reasonable; at luminosities above
$5 \times 10^{32}$~cm$^{-2}$
 sec$^{-1}$, all device occupancies were less than 20\% of
conservative allowances.  Radiation dose levels were low; in the case
of the highest radiation region of the Silicon Vertex Tracker (SVT),
dose rates were 20\% of the pre-ordained budget.

As can be seen from Fig.~\ref{dfig:8}, the detector data logging efficiency is
high, particularly at this early phase of operation.  Typically, BaBar
logs $\geq$  90\% of the delivered luminosity.  It is much too early to expect
physics results.  However, some plots are shown which are
representative of the process of tuning up and monitoring the
performance of BaBar.  This is an evolving process and much work remains
to be done.

\begin{figure}[htbp]
\begin{center}
{\hspace{1in}\epsfbox{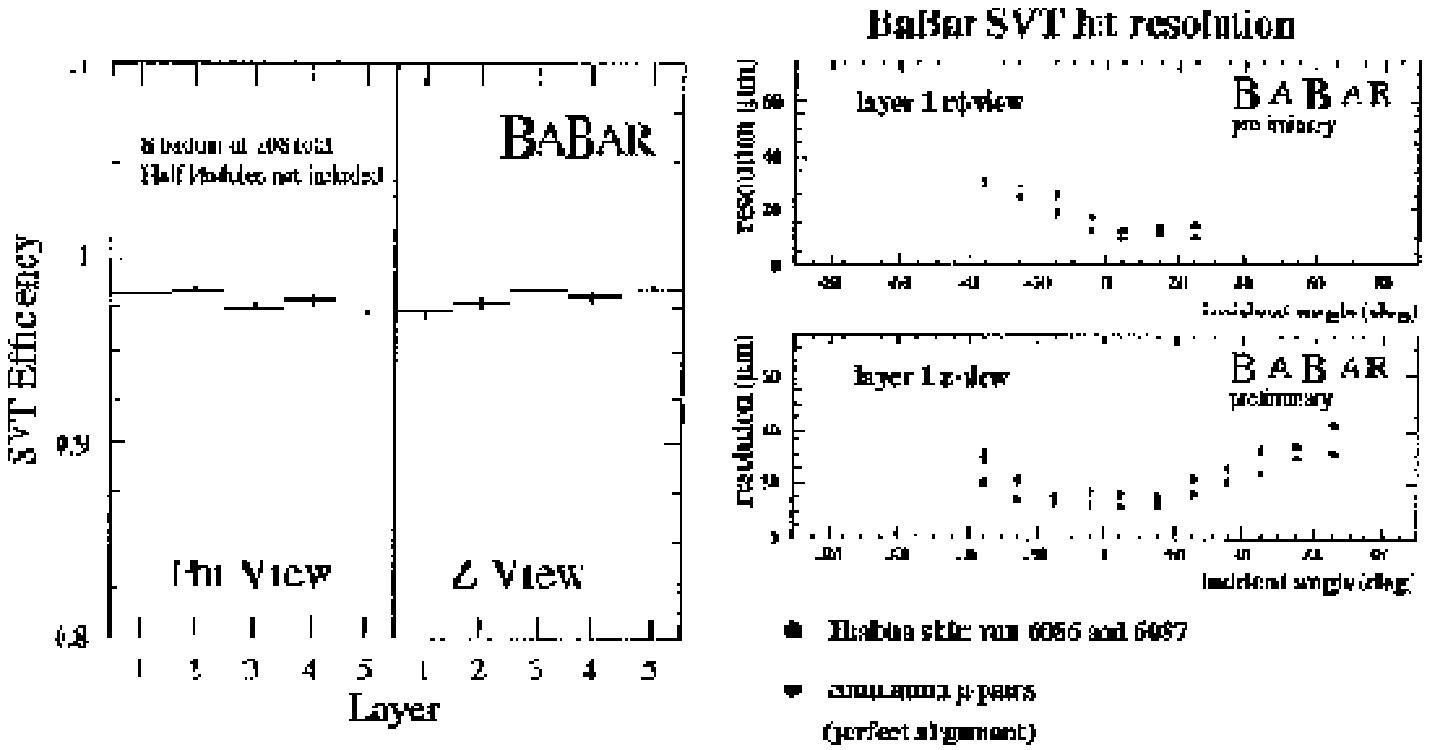}}
\caption[]{Performance of the BaBar silicon vertex detector.  Left: efficiency; Right:
tracking resolution.}
\label{dfig:9}
\end{center}
\vfill
\begin{center}
{\hspace{1in}\epsfbox{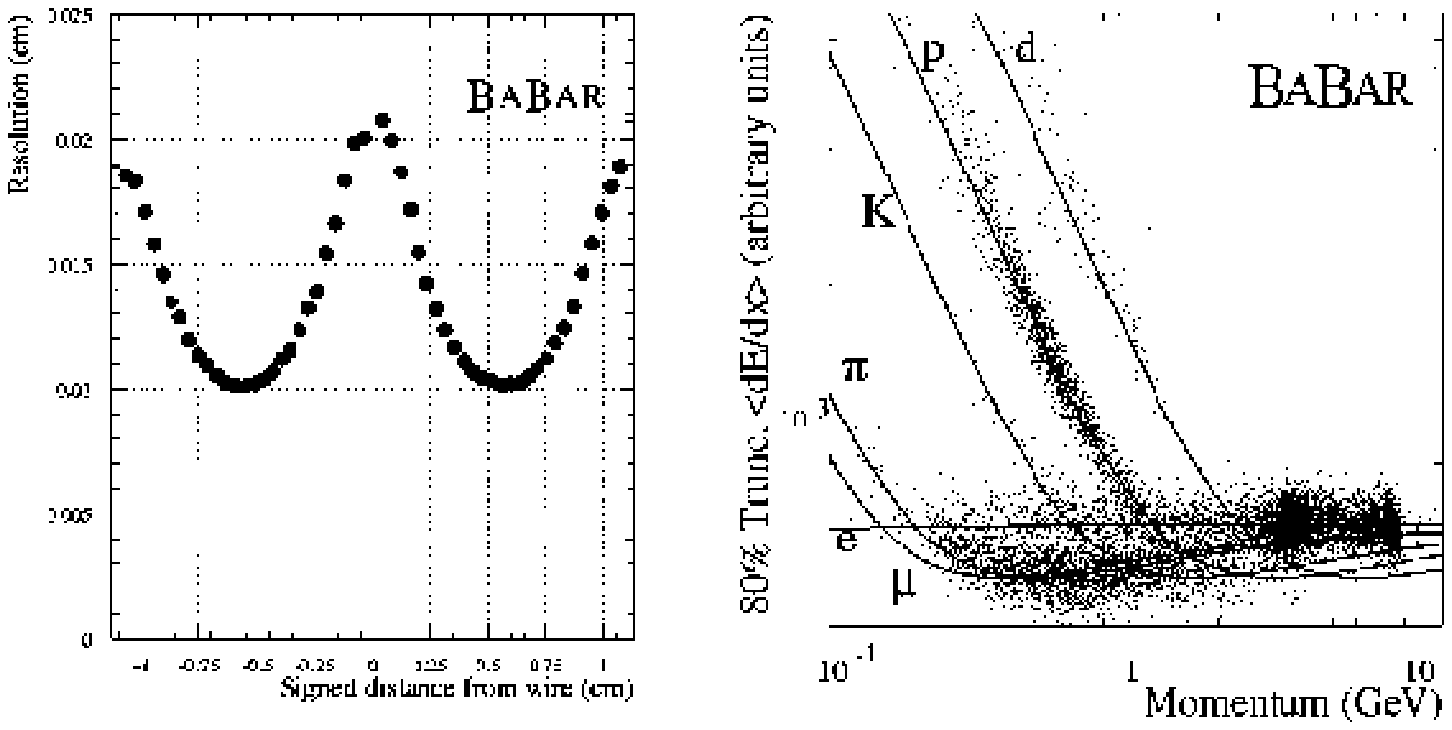}}
\caption[]{Performance of the BaBar drift chamber.  Left: tracking resolution;
Right: $dE/dx$.}
\label{dfig:10}
\end{center}
\end{figure}

Figure 9 shows early performance measures for the SVT.
If one excludes 8  non-working modules (out of 208), detector efficiencies are
seen to be excellent.  The hit resolution in $\phi$ and $z$ are not far from
that expected.

Figure~\ref{dfig:10} shows the drift chamber single-hit resolution as a function
of distance from the wire.  This resolution is very close to the design
performance of 140 $\mu$ averaged over the cell.
The current status of the $dE/dx$ performance is also shown in Fig.~\ref{dfig:10}.
The two dark spots at high momentum are the Bhabha electrons and positrons.  

\begin{figure}[htbp]
\begin{center}
{\hspace{1in}\epsfbox{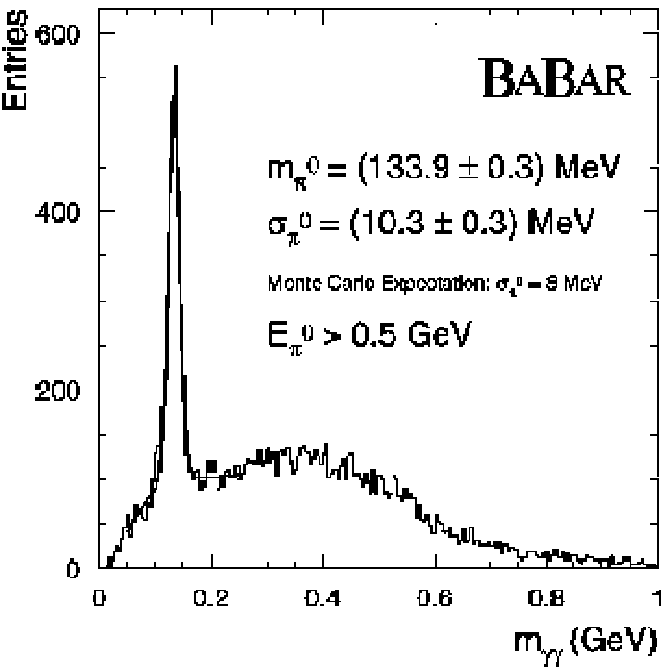}}
\caption[]{Two-photon mass spectrum observed by the BaBar detector. }
\label{dfig:11}
\end{center}
\vfill
\begin{center}
{\hspace{1in}\epsfbox{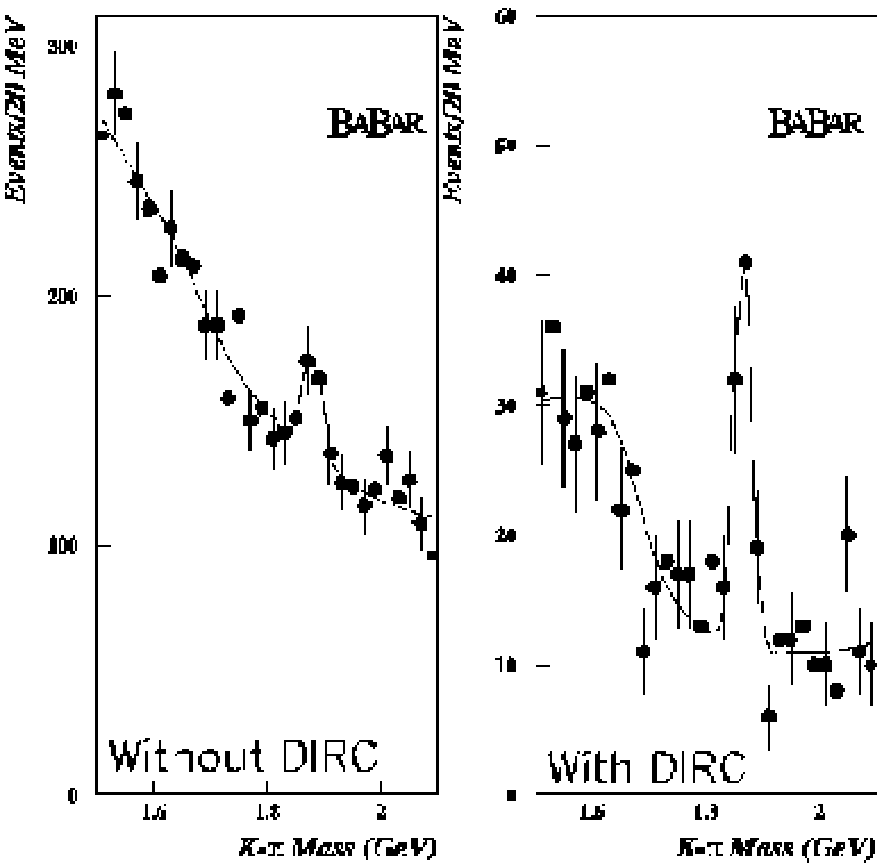}}
\caption[]{$K^\pm\pi^\mp$ mass spectrum observed by the BaBar detector.  Left: with
no particle ID; Right: including the DIRC system.}
\label{dfig:12}
\end{center}
\end{figure}

Figure~\ref{dfig:11}
shows the $\gamma\gamma$ mass spectrum for di-gamma energies above 500 MeV. A clear
$\pi^0$ peak is visible---the resolution on the $\pi^0$ is about 30\% larger than
design.

\begin{figure}[htbp]
\begin{center}
{\hspace{1in}\epsfbox{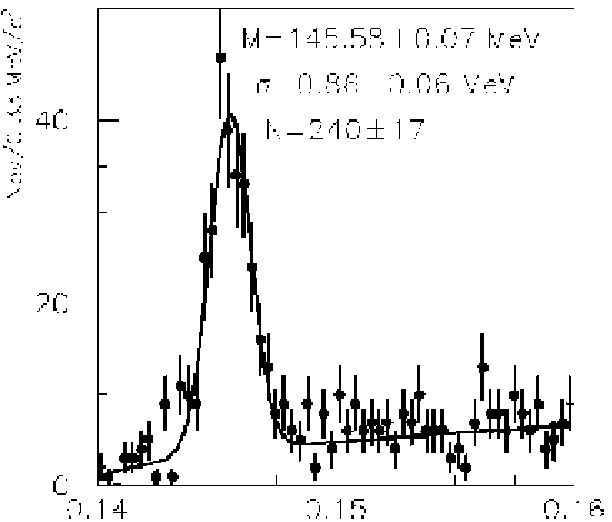}\hspace*{.1in} 
  \epsfbox{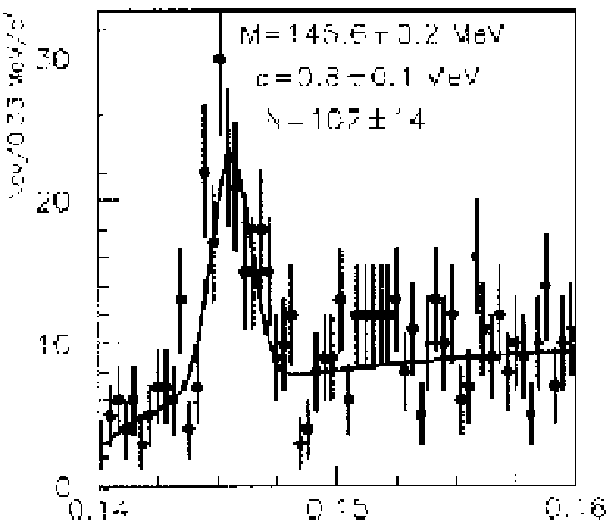}}
\caption[]{$D^*$-$D^0$ mass difference, as reconstructed by the BaBar detector.
Left: $D^0\rightarrow K^\pm\pi^\mp$; Right: $D^0\rightarrow K^\pm \pi^\mp\pi^0$.}
\label{dfig:13}
\end{center}
\vfill
\begin{center}
{\hspace{1in}\epsfbox{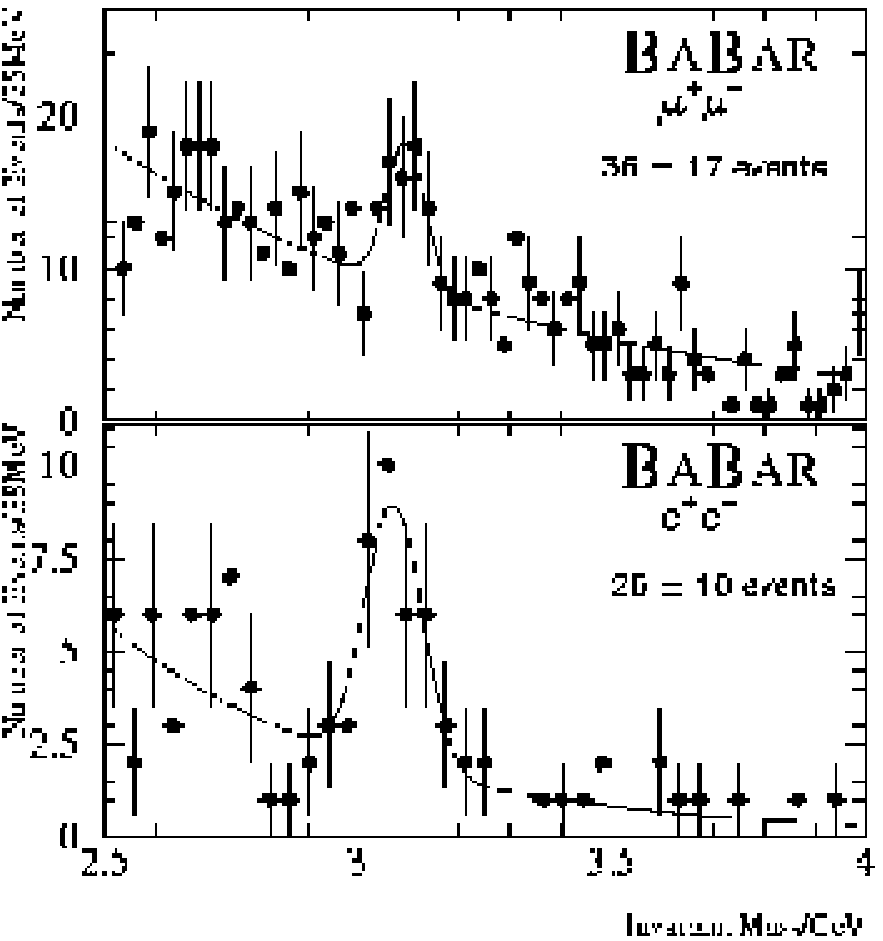}}
\caption[]{Dilepton invariant mass spectrum in BaBar events, showing the
$J/\psi \rightarrow \ell^+\ell^-$ peak.}
\label{dfig:14}
\end{center}
\end{figure}

Figure~\ref{dfig:12} shows the effect of \index{subject}{Cherenkov detector}
particle identification using the DIRC
system.  The $K^\pm\pi ^\mp$ mass spectrum in the region of the $D^0$ is shown without
using particle identification and using the DIRC to identify the kaon.

Figure~\ref{dfig:13} shows the $D^*$-$D^0$ mass difference, using the two $D^0$ decay modes,
 $K^\pm \pi^\mp$ and $K^\pm\pi^\mp\pi^0$ (no particle ID is used).  
These plots are indicative of good early performance of the tracking systems.

Figure~\ref{dfig:14} shows the di-lepton invariant mass spectra in the region of
the $J/\psi$ particle.

While it is much too early to expect high quality (design level)
performance from the detector, these plots demonstrate rather impressive
performance this early in the life of the detector.  

\section{Future Plans}

 As of LP99, the accelerator was complete, but BaBar was missing 2/3 of the
quartz bars.  These bars were installed in October 1999, thereby
completing the detector.  The facility will run until Summer 2000, with
a short break in December 1999.


\begin{thebibliography}{99}

\bibitem{ref1}
PEP-II, an Asymmtric B Factory: A Conceptual Design Report, 
LBL-PUB-5379, SLAC-R-418, CALT-68-1869,
UCRL-ID-114055, UC-IIRPA-93-01 (1993).

\bibitem{ref2}
D. Boutigny {\it et al.} [BaBar Collaboration] BaBar Technical Design Report,
SLAC-R-0457 (1995).


\end{thebibliography}
\end{document}